\documentclass[pra,twocolumn,10pt]{revtex4-2}

\usepackage{graphicx}
\usepackage{amsmath,bbm}
\usepackage{hyperref}
\usepackage{comment}
\usepackage{color}
\usepackage{appendix}
\usepackage{verbatim}
\usepackage{physics}
\usepackage{float}
\usepackage[normalem]{ulem}

\newcommand{\be}{\begin{equation}}
\newcommand{\ee}{\end{equation} }
\newcommand{\um}{\mathbbm{1}}

\graphicspath{{figures/}}
     
\begin{document}
\title{Robust Nonperturbative Trapped-Ion Quantum Logic}

\author{Luca Stefanescu, Florian Mintert}
\affiliation{Physics Department, Blackett Laboratory, Imperial College London, Prince Consort Road, SW7 2AZ, United Kingdom}

\begin{abstract}
Entangling gates of trapped ions are typically mediated by collective motional degrees of freedom.
Weak coupling between qubit and motional degrees of freedom and the resulting harmonic dynamics give access to a broad range of gate schemes, but also impose strict limitations on achievable gate times.
In this paper, we devise optimally designed driving schemes for the realization of fast, high-fidelity entangling gates mediated by anharmonic dynamics.
The driving can also be optimized to achieve resilience to multiple system imperfections, and the anharmonicity in the motional dynamics can be used to enhance such resilience.
\end{abstract} 
\maketitle

\section{Introduction}\label{sec:Introduction}

Trapped ions are among the best developed technologies for quantum information processing \cite{HAFFNER2008}.
Suspended in free space, qubits couple to the environment much more weakly than in competing technologies, such as superconducting qubits \cite{Kjaergaard2020}.
In many setups, the coherence time is thus limited by external quantities, such as the phase-stability of lasers to control quantum gates \cite{Schfer2018,Gaebler2016} or the stability of a magnetic field to lift degeneracies \cite{Wei2022}.

This weak intrinsic decoherence is an excellent precondition for the implementation of quantum algorithms of substantial depth.
However, in order to develop practical technologies that operate without an excessive effort of frequent system calibration, it is important that elementary gates have a healthy degree of resilience to external sources of decoherence \cite{Shapira2023,Le2025}.

Regarding the gate speed, trapped ions struggle in the comparison to superconducting qubits \cite{Kjaergaard2020}.
Given the mesoscopic separation between two trapped ions, there is no direct interaction worth mentioning, and effective interactions mediated by the strongly correlated real-space dynamics are required to realize entangling gates \cite{Cirac1995}.
The internal qubit degrees of freedom of the ions can be coupled to the motional degrees of freedom with driving in terms of electromagnetic fields.
Most existing gate schemes assume weak driving and a long wavelength of the driving field \cite{Mlmer1999,Srensen1999,Srensen2000}.
Sufficiently weak driving justifies taking into account only resonant or close-to-resonant transitions,
and a sufficiently long wavelength justifies treating the sinusoidal spatial dependence of the driving light field within a linear approximation.
In these limiting cases, the motional dynamics remains harmonic, and the simplicity of harmonic dynamics gives access to comparatively elementary control schemes in terms of monochromatic driving fields {\cite{Baldwin2021,Lishman2020,Leibfried2003,Zhang2025,Milburn2000,Roos2008,Schmidt-Kaler2004,Green2015,Leung2018,Choi2014,Zhu2006,Hayes2012}.

Driving schemes for the realization of fast gates thus nearly unavoidably require optimized driving patterns to ensure that the qubit-motion entanglement that is being established while the motion mediates an interaction vanishes at the gate time \cite{Anikin2025,Steane2014,Schfer2018,Palmero2017}.
In optimizing driving patterns, one may also ensure that quantum gates have a given degree of noise resilience.
Driving fields can be designed to make quantum gates resilient to fluctuations in system parameters, such as a magnetic field that determines the qubit resonance frequencies \cite{Lishman2020}, the trap frequency \cite{Liu2025}, or the Rabi frequency \cite{Shapira2023,Le2025}.
Since achieving such resilience necessarily exploits non-commutativity, it can be beneficial if the spatial dynamics is not perfectly linear because the nonlinear dynamics gives access to a broader class of time-evolution operators with greater potential to find terms with suitable commutation relations.

While initially considered an unavoidable problem to address in the quest for higher gate speeds, a nonlinearity thus also has beneficial properties for the realization of practical gates.
The goal of this paper is thus to design control schemes for trapped ions that admit strong driving and that induce fast and noise-resilient entangling gates.

\section{Driven trapped ions}

A system of two trapped ions is characterized by the Hamiltonian
\be
H_0=
\frac{\omega_0}{2}\sum_{j=1}^2\sigma_z^j+
\sum_{j=1}^2\nu_ja_j^\dagger a_j\,,
\ee
where $\sigma_z^j$ is the Pauli Z matrix for ion $j$;
$\nu_j$ are the eigenfrequencies of the two collective modes of oscillation along the axis of weak confinement of the trapping potential; $a_j$ and $a_j^\dagger$ are the corresponding annihilation and creation operators.
For the center of mass (COM) and relative motion the eigenfrequencies are given by $\nu_1=\omega_T$ and $\nu_2=\sqrt{3}\omega_T$ in terms of the angular trap frequency $\omega_T$ with the corresponding COM trap period $T=2\pi/\omega_T$.

The interaction of an ion with an in-phase electromagnetic driving field with frequency $\omega$ is captured by the interaction Hamiltonian
\be
H_c
=\Omega \sigma_x\cos(kx-\omega t)\,,
\label{eq:elementaryinteration}
\ee
where $x$ is the displacement operator of the driven ion, $k$ is the component of the driving field's wave-vector along the trap axis, $\Omega$ is the Rabi frequency of the driving field.
The interaction with a phase-shifted driving field reads
\be
H_s
=\Omega \sigma_x\sin(kx-\omega t)\,.
\ee

In the specific case of two ions, 
the displacement of each of the two ions is expressed in terms of the normal mode operators via the relation
\be
kx_j=\eta\left( (a_1+a_1^\dagger)-\frac{(-1)^j}{3^{1/4}} (a_2+a_2^\dagger)\right)\,,
\label{eq:kx}
\ee
with the Lamb-Dicke (LD) parameter
\be
\eta=k\sqrt{\frac{\hbar}{2\mu\omega_T}}\,,
\ee
in terms of the mass $\mu$ of an ion.

\subsection{Driving scheme}

In the design of a driving scheme for the implementation of an entangling gate, there is a tradeoff between what can be achieved with a large number of driving fields and the resultant experimental complexity.
The following analysis is based on two in-phase, counter-propagating driving fields and two additional phase-shifted counter-propagating driving fields.
The complete interaction Hamiltonian is thus of the form

\be
\sum_{j=1}^{2}\sigma_x^j
\sum_{q=\pm 1}\left[
\Omega_1\cos(qkx_j-\omega_1t) - q
\Omega_2\sin(qkx_j-\omega_2t)\right]\,,
\label{eq:interaction}
\ee
with the Rabi frequencies of the in-phase and phase-shifted driving fields denoted by $\Omega_1$ and $\Omega_2$, and a relative $\pi$ phase shift between the two directions of the terms with Rabi frequency $\Omega_2$.
Ideally, all the driving fields would be on resonance with the qubit transitions, i.e., $\omega_1=\omega_2=\omega_0$, but Eq.~\eqref{eq:interaction} takes into account deviations from perfectly resonant driving.

Common approximations invoked in the design of driving fields for quantum gates involve restricting the Hamiltonian to one (close-to)-resonant motional mode and assuming sufficiently small displacements from the ions' rest positions such that the spatial dependence in the interaction (Eq.~\eqref{eq:elementaryinteration}) can be approximated as a low-order power-series in the LD parameter $\eta$.

While these approximations may be sufficiently good for proof-of-principle demonstrations of entangling gates or for slow gates, their accuracy can become insufficient for the realization of gates with the high fidelities required for quantum error correction even at moderate coupling or driving strength.
In particular, with increasing driving strength required for fast gates, these approximations break down \cite{Schfer2018,Ballance2016,Gaebler2016}, and modeling in terms of more accurate approximations is required.

The subsequent analysis considers driving the qubits on or close to resonance, 
and transitions in the system Hamiltonian
that are off-resonant by at least the qubit resonance frequency $\omega_0$ are neglected.
The full system Hamiltonian in this rotating wave approximation is given in Eq.~\eqref{eq:fullH} in the appendix.
If the two driving frequencies $\omega_1$ and $\omega_2$ coincide,
the system Hamiltonian simplifies;
in the frame defined by the free dynamics of the qubit Hamiltonian $\frac{\omega_0}{2}\sum_{j=1}^2\sigma_z^j$ it reads
\begin{equation}
    H = \sum_{j=1}^2\nu_ja_j^\dagger a_j + \left[ \sum_{j=1}^2\left( \sigma_+^j e^{-i \Delta t} + \sigma_-^j e^{i \Delta t} \right)
\Omega_R e^{ikx_j} + \mathrm{h.c.}\right],
\label{eq:Hfinal}
\end{equation}
in terms of the detuning $\Delta=\omega_1-\omega_0=\omega_2-\omega_0$, with the dependence on the displacements $x_j$ of the two ions expressed in terms of Eq.~\eqref{eq:kx},
and with the complex Rabi frequency
\be
\Omega_R(t)=\Omega_1(t)+i\Omega_2(t)\,.
\ee

The goal of the following analysis is the identification of time-dependent patterns for the Rabi frequency $\Omega_R(t)$ for the realization of fast and noise-resilient high-fidelity gates.
In Sec.~\ref{sec:noise} this scope will be expanded to also include fluctuations in the Rabi frequencies and the detunings of the driving fields.

\subsection{Optimal control}

Given the anharmonic system dynamics, there is no realistic prospect of identifying suitable driving schemes analytically, but the system Hamiltonian is amenable to a numerically exact treatment with the motional dynamics captured by a truncated Fock space.
Any dependence of a gate on the initial motional state or fluctuations of Rabi frequency and laser detuning can be actively taken into account via the optimization target.

The system Hamiltonian inherently implies dynamics in which qubit and motional degrees of freedom are interacting with each other.
Instead of a unitary of only the qubit degrees of freedom, the current problem thus needs to be formulated in terms of a target unitary on all involved degrees of freedom.
Given that an ideal entangling gate should be completely independent of the initial motional state, the control target would ideally read
$U_T=U_Q\otimes\um_M$ with $U_Q$ being the desired entangling gate of the qubit degrees of freedom and $\um_M$ being the identity on the motional degrees of freedom.
Since, in practice, it is not possible to realize an entangling gate that is completely independent of the initial motional states, the control target for the subsequent discussion reads
\be
U_T=U_Q\otimes P_M\,,
\label{eq:Utarget}
\ee
where $P_M$ is the projector onto the subspace of the motional Hilbert space spanned by the initial states for which close-to-perfect gates are desired.
An optimal control solution obtained with the target $U_T$ (Eq.~\eqref{eq:Utarget}) realizes an entangling gate that is independent of the motional initial state, as long as this state is within the subspace defined by the projector $P_M$.

In order to assess the accuracy of gates realized with optimized driving,
it is helpful to define the thermal infidelity
\be
I_\beta(V)=1-\sum_{\bf n} P_\beta({\bf n})F(V|{\bf n})\,,
\label{eq:Ithermal}
\ee
with the Boltzmann weight factor
\be
P_\beta({\bf n})=\prod_{j=1}^{2}
e^{-\beta\nu_jn_j}
(1-e^{-\beta\nu_j})\,,
\ee
and the gate fidelity
\be
F(V|{\bf n})=\frac{1}{d^2}
\sum_{ijm}
\bra{\Phi_i}U_Q^\dagger
K_{{\bf nm}}
\ket{\Phi_i}\bra{\Phi_j}
K_{{\bf nm}}^\dagger
U_Q\ket{\Phi_j}\,,
\label{eq:gatefidelity}\ee
obtained given an initial motional state $|{\bf n}\rangle=|n_1,n_2\rangle$ and a $d$-dimensional computational Hilbert space,
where the state vectors $\ket{\Phi_i}$ form an orthonormal basis for the qubit degrees of freedom,
and with the Kraus operators
\be
K_{nm}=\bra{{\bf m}}V\ket{{\bf n}}\,,
\ee
in terms of a complete set of motional states $\ket{{\bf m}}$ that capture the reduced qubit dynamics resulting from the unitary dynamics $V$ of the complete system given the initial motional state $\ket{{\bf n}}$.

Fluctuations in classical parameters like Rabi frequency or laser detuning can be taken into account via the method of ensemble control \cite{Li2009}, where an ensemble average over system dynamics $V_i$ is used, such that
each propagator $V_i$ is induced by a Hamiltonian $H_i$ with a randomly chosen value for each of the fluctuating parameters.
This yields the ensemble-averaged infidelity
\be
\bar I=1-\langle F(V_i|{\bf n})\rangle_{i,{\bf n}}\,.
\label{eq:ensemblefid}
\ee

\section{Optimized entangling gates}

The system Hamiltonian $H$ (Eq.~\eqref{eq:Hfinal}) and the gate fidelity with the control target $U_T$ (Eq.~\eqref{eq:Utarget})
define an optimal control problem for the design of a time-dependent Rabi frequency $\Omega_R(t)$.
All of the results presented in the following rely on numerically optimized time-dependence designed to achieve the target propagator $U_Q=\exp(i\frac{\pi}{4}\sigma_x^1\otimes\sigma_x^2)$
that allows for the creation of maximally entangled states given initial separable states.

Fig.~\ref{fig:speed} depicts the gate infidelity as a function of the gate duration $T_G$ for different values of $\eta$.
The driving patterns are optimized for the control target $U_T$ (Eq.~\eqref{eq:Utarget}) where $P_M$ is the projector onto the space spanned by the two Fock states $|0,0\rangle$ and $|1,0\rangle$.
To ensure consistency across all figures, the explicit form of the infidelity depicted in Fig.~\ref{fig:speed} is the average gate infidelity $1-\sum_{{\bf n}}F(V|{\bf n})/2$ with $F(V|{\bf n})$ defined in Eq.~\eqref{eq:gatefidelity} and where the summation includes the same two initial Fock states that were used in the optimization.

\begin{figure}
\centering
\includegraphics[width=\columnwidth]{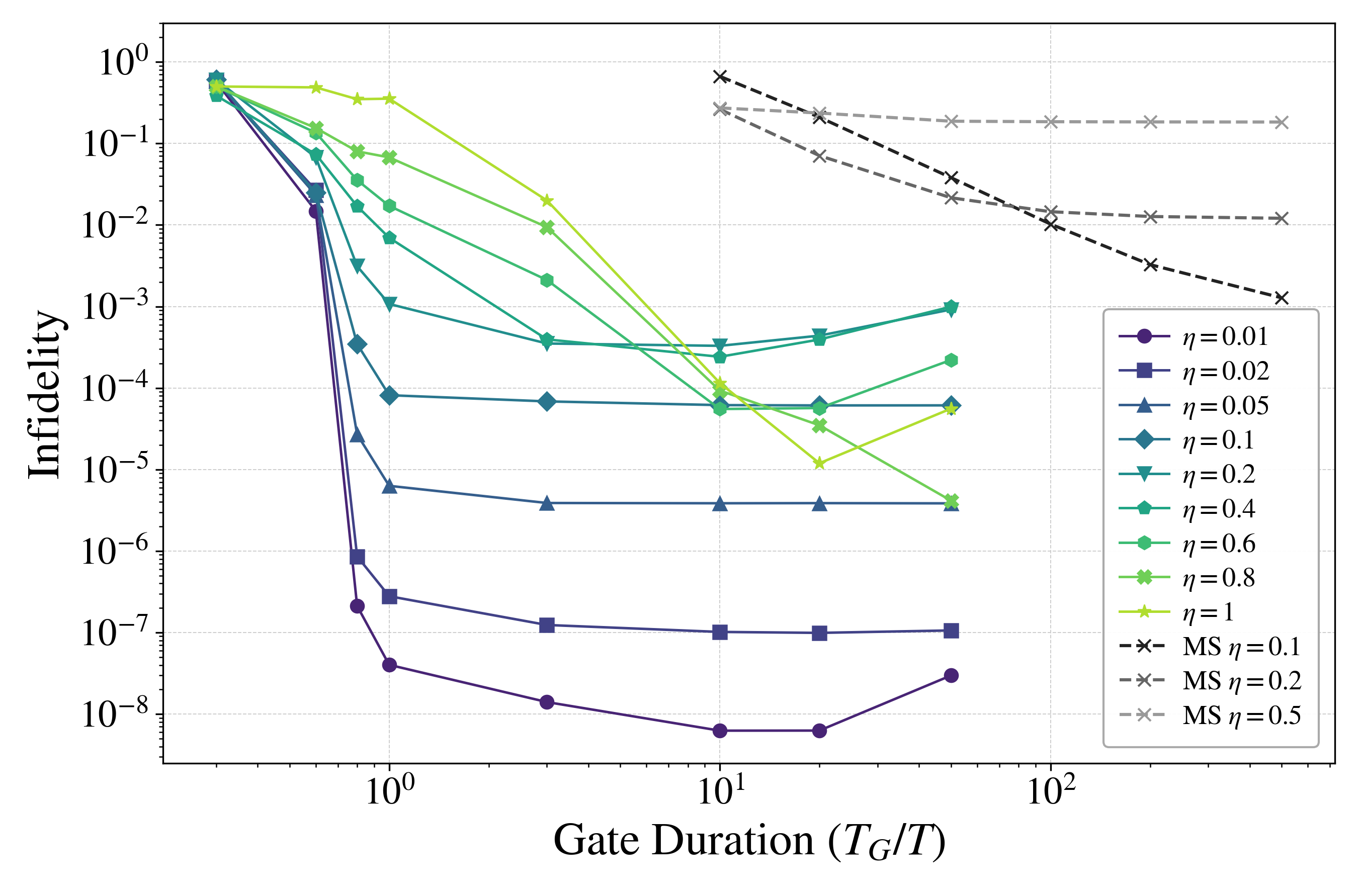}
\caption{Infidelity of entangling gates resulting from optimized driving as a function of dimensionless gate duration $T_G/T$ and the LD parameter $\eta$.
For sufficiently small values of the LD parameter $\eta$, there is a pronounced drop in infidelities for gate durations exceeding one trap period $T$ and a nearly constant infidelity for longer durations.
For larger values of $\eta$ this clear step-like behavior gets washed out, and the increasingly nonlinear behavior of the system dynamics requires longer gate durations for high-fidelity entangling gates.
Infidelities of gates realized with M\o lmer-S\o rensen (MS) driving (Eq.~\eqref{eq:MS}) are depicted for comparison for gate times exceeding $10$ trap periods $T$.
The lines connecting data points are a guide to the eye.}
\label{fig:speed}
\end{figure}

Since the entangling interaction is mediated by the collective motion of the ions
with a natural timescale given by
the trap period,
this timescale is a fundamental limitation to the achievable gate times.
This is reflected in the sharp rise in infidelity for gate times shorter than the trap period $T$.
Gates over a duration of one trap period deliver high fidelities in the case of sufficiently small values of the LD parameter $\eta$.
In fact, for $\eta\le 0.2$ there is only a close-to-negligible reduction in gate infidelity with an increase in gate time.

Only for larger values of the LD parameter ($\eta\ge 0.4$) is there a substantial decrease of gate infidelity with increasing gate time.
This can be attributed to the highly anharmonic character of the spatial dynamics with high values of the LD parameter.
This dynamics makes the motional quantum state of the ions adopt strong non-Gaussian characteristics as the driving sets in, and it takes some time for the controlled dynamics to have the motional state evolve towards the initial state as required for high-fidelity, coherent gate operation.

\subsection{Comparison to common gate scheme}

\begin{figure}
\centering    \includegraphics[width=\columnwidth]{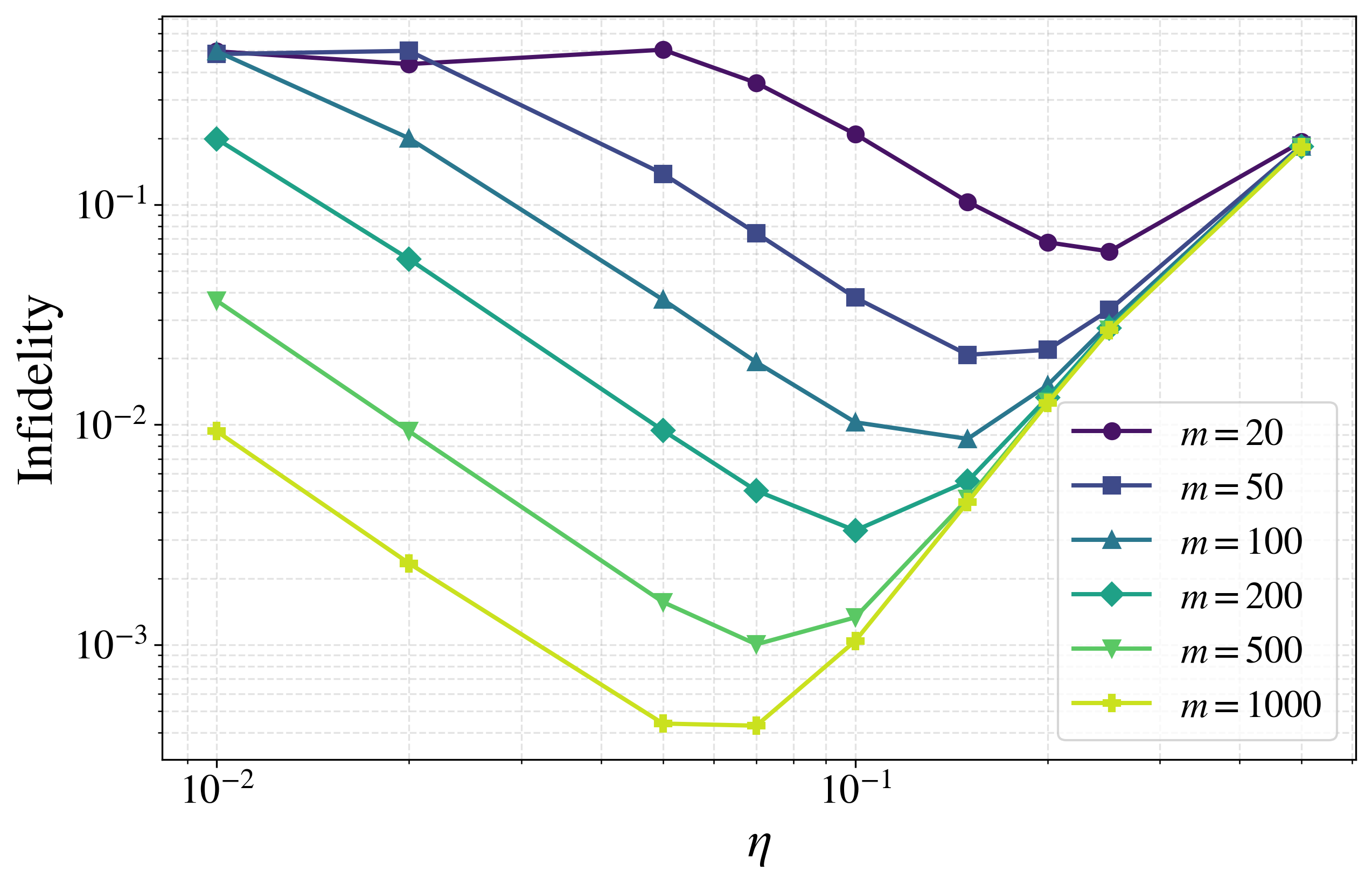}
\caption{Infidelities of entangling gates induced by Eq.~\eqref{eq:HMS} with the regular MS drive as a function of the LD parameter $\eta$ and the dimensionless gate duration $T_G/T=m$ as a multiple of the trap period $T$. This demonstrates that each duration corresponds with a specific optimal $\eta$, and this optimal point is lower when gate duration is larger. Below this optimal $\eta$, the infidelity increases due to off-resonant effects. The lines connecting data points are a guide to the eye.}
\label{fig:ms}
\end{figure}

The benefits of the optimized driving can be assessed in a comparison with the gate fidelities obtained with conventional driving schemes for system dynamics induced by the system Hamiltonian without the approximations that conventional driving schemes rely on.
The MS scheme \cite{Srensen1999,Srensen2000,Mlmer1999} is a natural choice for such a comparison, but it is not directly applicable to the system Hamiltonian $H$ in Eq.~\eqref{eq:Hfinal}, since this Hamiltonian is based on pairs of counter-propagating driving fields, 
whereas the MS scheme can be implemented with fields propagating in just one spatial direction.

In the absence of the counter-propagating fields, the Hamiltonian $H$ in Eq.~\eqref{eq:Hfinal} reduces to
\begin{equation}
H_{MS} = \sum_{j=1}^2\nu_ja_j^\dagger a_j + \left[\sum_{j=1}^2
\Omega_{MS}
\sigma_+^j e^{-i \Delta t}
 e^{ikx_j} + \mathrm{h.c.}\right]\,,
\label{eq:HMS}
\end{equation}
with the analytically derived MS drive given by
\begin{align}
\Omega_{MS} &= \frac{\delta}{2\eta}\sin(\delta t - \nu_1 t)\,,
\label{eq:MS}
\end{align}
where the detuning from the COM mode $\nu_1$ is given by $\delta=\frac{2\pi}{T_{MS}}\ll\nu_1$ and $T_{MS}$ is the MS gate duration.

Gate fidelities resulting from the dynamics induced by $H_{MS}$ with the driving $\Omega_{MS}$ are depicted in Fig.~\ref{fig:speed} for gate times exceeding $10T$.
Even a gate time of $10T$ is substantially too short to achieve well-functioning gates.
The gate infidelities do decrease with increasing gate time, but a value of the LD parameter not larger than $0.1$ is required to avoid saturation of the gate infidelity with gate times exceeding $200T$.

Even with a small value of the LD parameter (e.g., $\eta=0.1$), which allows for high fidelities of slow  quantum gates, it is not possible to increase the gate speed with stronger driving without reducing the gate fidelity.
This aspect can also be seen in Fig.~\ref{fig:ms}, which depicts the gate infidelity obtained with the driving pattern Eq.~\eqref{eq:MS} for different gate durations as a function of the LD parameter $\eta$.
For any given gate duration, there is an optimal value of $\eta$; its value lies around $0.1$ for slow gates and it increases with decreasing gate durations.

Irrespective of the gate duration, it is not possible to arrive at a perfect gate through the limit $\eta\to 0$.
This reflects the fact that the driving in Eq.~\eqref{eq:MS} is effective only if both the LD approximation ($\eta\ll 1$) and the weak driving approximation that takes into account only close-to-resonantly driven transitions are valid.
The carrier transitions are driven with an amplitude $\Omega_R$ and a detuning of the order of the trap frequency $\omega_T$, while the close-to-resonantly driven sideband transitions are driven with an amplitude $\eta\Omega_R$.
In the regime of validity of the LD approximation, the carrier transition is thus driven more strongly than the sideband transitions, and correspondingly weak driving is required such that the off-resonant nature of the carrier transitions is enough to neglect their impact on the gate.

Since the gates realized with optimized driving do not require neglecting contributions of off-resonantly driven transitions, this limitation does not apply, and it is indeed possible to identify driving patterns to achieve gates with durations that are limited only by the trap period.

\subsection{Driving amplitude}

\begin{figure}
    \centering
    \includegraphics[width=\columnwidth]{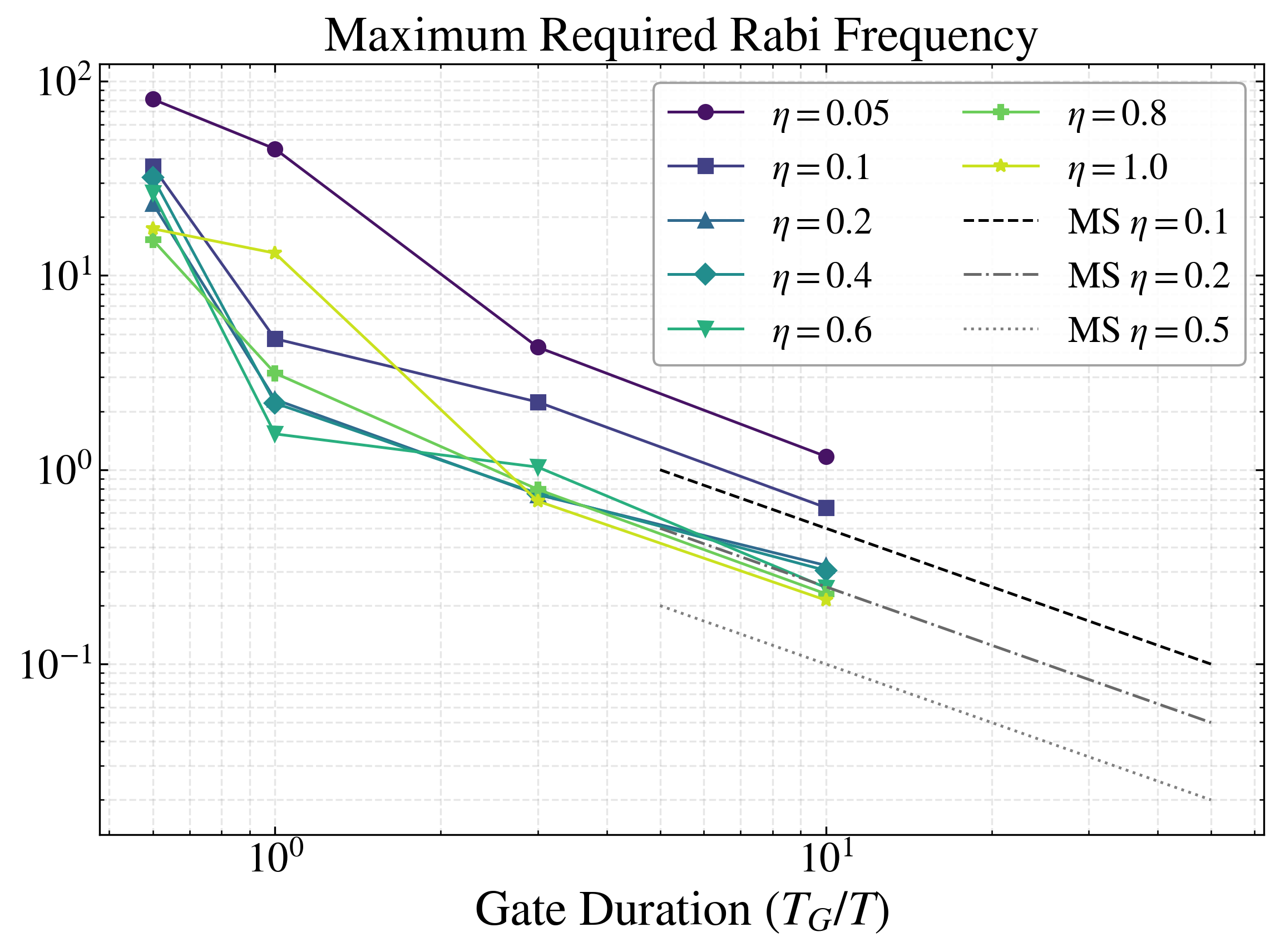}
    \caption{Dimensionless maximal value of the Rabi frequency ($\max_{i,t}(|\Omega_i(t)|/\nu_1)$) required by the optimized drives (colored lines) and by the MS scheme (dashed lines) as a function of dimensionless gate duration and the LD parameter $\eta$. The dependence of the amplitude of the optimized drives on the duration is qualitatively the same as for the MS drives. The colored lines connecting data points are a guide to the eye, the dashed lines show the analytic dependence.}
    \label{fig:ampl}
\end{figure}

As seen in Fig.~\ref{fig:speed}, the optimized driving can help to overcome the limitations caused by off-resonantly driven transitions and by anharmonic dynamics that apply to analytically derived driving patterns.
Fundamentally, however, fast gates require strong driving, and limitations
to realizable driving strength can become a relevant impediment \cite{Ge2019}.

Fig.~\ref{fig:ampl} depicts the dimensionless maximal value ($\max_{i,t}(|\Omega_i(t)|/\nu_1)$) of the Rabi frequencies required to drive gates as a function of their duration.
Similarly to Fig.~\ref{fig:speed}, this is shown for the optimized driving patterns and, for longer gate times, also for the driving $\Omega_{MS}$.
The MS driving has the analytic dependence $|\Omega_{MS}|\propto \frac{1}{\eta T_{MS}}$.
The optimized driving patterns show qualitatively the same dependence (this can also be seen in Fig.~\ref{fig:ampleta}), and the deviation from the exact dependence can well be attributed to the non-uniqueness of optimized solutions that can be found with numerical optimizations.
While the optimized solutions achieve the tasks of compensating for the impact of a broad range of processes that would reduce the gate fidelity with elementary driving patterns, this does not come with an increase in the required driving strength beyond what one would have expected from the scaling of existing gate schemes.

\subsection{Resilience to different noise sources}
\label{sec:noise}

\begin{figure*}[t]
\centering
\includegraphics[width=\textwidth]{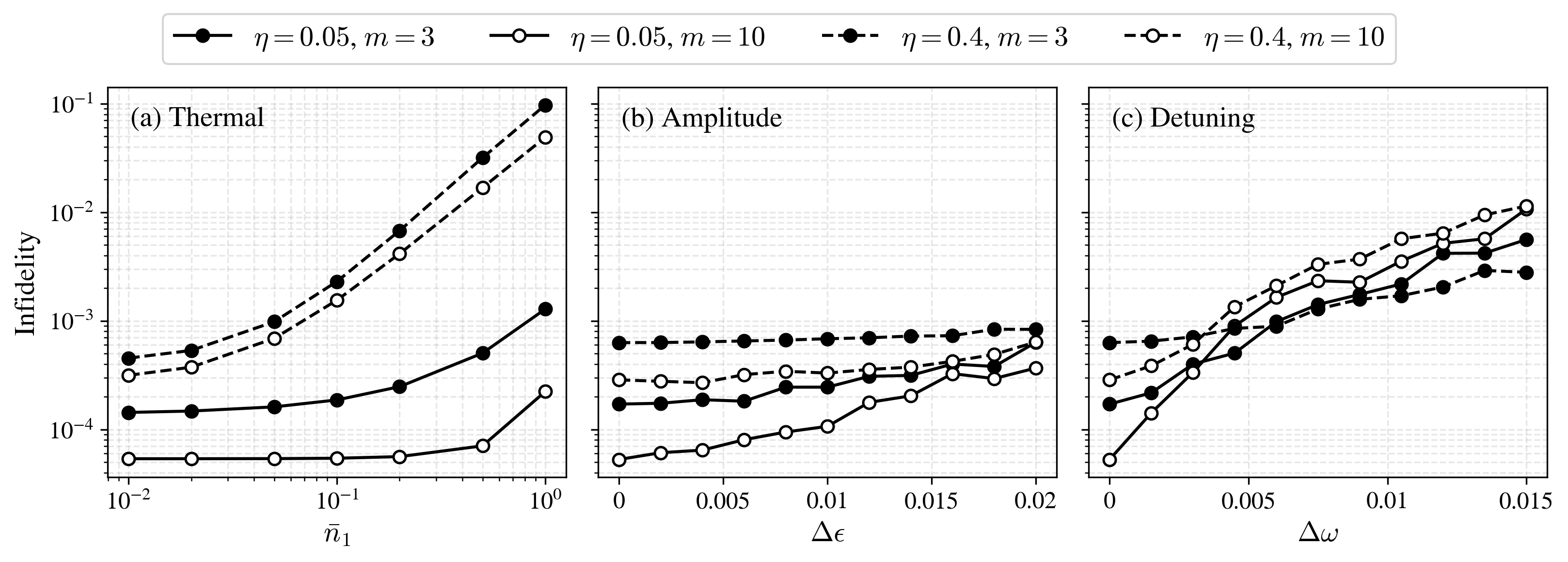}
    \caption{(a) Thermal infidelity of gates as defined in Eq.~\eqref{eq:Ithermal} with $\Delta\epsilon=0$ and $\Delta\omega=0$ as a function of the mean thermal occupation $\bar{n}_1$. The infidelities stay below $10^{-4}$ when weak coupling and large duration are used, but they increase faster as temperature increases when strong coupling is used. (b) Ensemble-averaged infidelity (Eq.~\eqref{eq:ensemblefid}) as a function of laser intensity noise level $\Delta\epsilon$, with $\bar{n}_1=0$ and $\Delta\omega=0$. Infidelities remain below $10^{-3}$ for all durations and values of $\eta$, across all noise levels tested. (c) Ensemble-averaged infidelity as a function of detuning noise level $\Delta\omega$, with $\bar{n}_1=0$ and $\Delta\epsilon=0$. In the noiseless case the slow and weak coupling gate achieves the lowest infidelity, but when the noise level is increased, the faster gate with larger $\eta$ achieves the lowest infidelity. Results are shown for gate durations of 3 and 10 trap periods, and LD parameter values of $0.05$ and $0.4$. $\epsilon_{ij}$ and $\varepsilon_{ij}$ as defined in Eqs.~\eqref{eq:Wran} and \eqref{eq:wran} are randomly drawn from uniform distributions with the ranges $[-\Delta\epsilon,\Delta\epsilon]$ and $[-\Delta\omega,\Delta\omega]$ respectively. The lines connecting data points are a guide to the eye.}
    \label{fig:allresilience}
\end{figure*}

Given the ability to drive high-fidelity gates despite nonlinear motional dynamics with moderately intense driving fields, the sensitivity of quantum gates to the temperature and to fluctuations in the driving patterns is the most relevant limitation to practically achievable gate fidelities.
It is thus desirable to include resilience to fluctuations in these quantities as additional targets of optimization.

To this end, the following discussion is based on an ensemble of system Hamiltonians $H_i$ (Eq.~\eqref{eq:Hfinal}) such that each ensemble member has Rabi frequencies $\Omega_{ij}$ and detunings $\Delta_{ij}$ which read
\begin{eqnarray}
\Omega_{ij}&=&(1+\epsilon_{ij})\Omega_j^c\,,
\label{eq:Wran}\\
\Delta_{ij}&=&\Delta^c+\varepsilon_{ij}\omega_T\,,
\label{eq:wran}
\end{eqnarray}
in terms of the central Rabi frequency $\Omega_{i}^c$ and the noise-less detuning $\Delta^c$,
and with random numbers $\epsilon_{ij}$ and $\varepsilon_{ij}$ drawn from a uniform distribution.
The aim of the optimization is the construction of time-dependent central Rabi frequencies $\Omega_i^c$.
The temporal profiles for the subsequently discussed cases are obtained from an optimization with an ensemble of 10 Hamiltonians (Eq.~\eqref{eq:fullHnoise}) with randomly chosen values of $\epsilon_{ij}$ and $\varepsilon_{ij}$ distributed uniformly within the range $[-0.01,0.01]$.
The accuracy of the gates resulting from those driving patterns is assessed with ensemble-averaged infidelities (Eq.~\eqref{eq:ensemblefid}) based on separate test ensembles in which $\epsilon_i$ and $\varepsilon_i$ are distributed within $[-\Delta\epsilon,\Delta\epsilon]$ and $[-\Delta\omega,\Delta\omega]$ respectively.
In particular, in order to highlight the dependence of the infidelity of the gates obtained with the optimized driving patterns on the noise-level, the ensemble-averaged infidelities will be discussed as a function of the range of these test ensembles ($\Delta\epsilon$ and $\Delta\omega$).

The numerical simulations presented here do not require the definition of an actual timescale or frequency scale, but all durations and frequencies can be specified in multiples of the trap frequency or its inverse.
For the sake of comparison to actual experiments, however, it can be helpful to recall that for trap frequencies of the order of a few MHz and gate durations of 1 to 50 trap periods, 
the fluctuations in the detuning
considered here are of the order of tens of kHz and 
the fluctuations in the Rabi frequency 
are of the order of hundreds of kHz.

Fig.~\ref{fig:allresilience} depicts average infidelities
of gates resulting from time-dependent Rabi frequencies optimized with an ensemble that includes fluctuations on both the Rabi frequency and the detuning, and an objective function projector specifying the two lowest energy Fock states (Eq.~\eqref{eq:Utarget} with $P_M$ projecting to $|0,0\rangle$ and $|1,0\rangle$).
Each of the three panels depicts gates in four different parameter regimes.
The infidelities for the faster gates ($T_G=3T$) are depicted with filled points, while the slow ones ($T_G=10T$) are depicted with hollow points. The infidelities with $\eta=0.05$ are depicted with a solid line while the ones with $\eta=0.4$ are depicted with a dashed line.

Fig.~\ref{fig:allresilience}a depicts thermal infidelities (Eq.~\eqref{eq:Ithermal}) with $\Delta\omega=0$ and $\Delta\epsilon=0$ as a function of the thermal occupation $\bar n_1$ of the COM mode (a proxy for  temperature) of the initial motional state.
As one would expect, infidelities are lowest at lower temperatures, longer durations, and weaker coupling.
In the two cases with small value of the LD parameter, the increase in infidelity with increasing temperature is slow compared to the cases with higher value of the LD parameter.
The faster gates have slightly higher infidelities than the slower gates, but even the fast gates with a duration that is close to the minimal required gate duration as identified in Fig.~\ref{fig:speed} have infidelities on the order of $10^{-4}$.
Gates in systems with strongly nonlinear qubit-motion coupling 
also have low infidelities (lower than $10^{-3}$) for sufficiently low temperature.
The increase in gate infidelity with increasing temperature is however more pronounced than in the cases with weaker, close-to-linear qubit-motion coupling.
Given the complete independence from thermal excitations that can be achieved in the LD regime of perfectly linear qubit-motion coupling, this behavior is not necessarily unexpected, but it hints at a tradeoff of strong qubit-motion coupling:
it enables the realization of fast gates with comparatively weak driving, but increasing the value of $\eta$ implies more stringent requirements on the cooling that is necessary for the implementation of high-fidelity gates. 

Fig.~\ref{fig:allresilience}b depicts the ensemble-averaged gate infidelity (Eq.~\eqref{eq:ensemblefid}) for test-ensembles with $\Delta\omega=0$ as a function of $\Delta\epsilon$.
Even though the control pulses are designed based on an ensemble with an interval $[-0.01,0.01]$, there is no clearly pronounced increase in the infidelity as the width of the test-ensemble exceeds the width of the ensemble used to construct the control fields.
That is, even if in an experimental implementation the fluctuations in the Rabi frequency are larger than assumed, one would expect to still obtain high-fidelity gates.
Similarly to the case of thermal excitations discussed in Fig.~\ref{fig:allresilience}a, the resilience to Rabi frequency noise is also most pronounced for small values of $\eta$ and for slow gates.
This improved performance is, however, most pronounced in the regime of small fluctuations in the Rabi frequency.
For fluctuations in the Rabi frequency exceeding $1\%$, the gap between the different infidelity curves decreases, such that gates in the challenging parameter regimes of strong coupling and short gate durations perform nearly as well as slower gates with close-to-linear qubit-motion coupling.

The increased resilience to fluctuations in the Rabi frequency for small values of $\eta$ observed in Fig.~\ref{fig:allresilience}b seems surprising, since gate schemes based on spectrally resolved driven transitions require a sizable nonlinearity (such as a sufficiently large value of $\eta$) in order to be resilient to fluctuations in the Rabi frequency~\cite{Shapira2023,Le2025}.
Since the present driving schemes do not aim at merely driving some selected sideband-transitions that realize an effective entangling interaction,
but they unavoidably also drive carrier transitions of the ions, the driven dynamics induces some additional non-commutativity in the propagator even for small values of $\eta$ \cite{Steane2014,Palmero2017}, which is not present in conventional schemes.
As such, the present approach can help to realize gates that are resilient to Rabi frequency fluctuations, even in the limit of very weak qubit-motion coupling.

The final aspect addressed in Fig.~\ref{fig:allresilience} is the dependence of the gate infidelities on fluctuations in the detuning.
Fig.~\ref{fig:allresilience}c depicts the infidelity (Eq.~\eqref{eq:ensemblefid}) as a function of $\Delta\omega$ for $\Delta\epsilon=0$.
Only in the close-to-noiseless case do the slow gates for the smaller value of $\eta$ yield the lowest infidelities.
In fact, for noise in the detuning above $1\%$, the fast gate of the system with the stronger coupling yields the lowest infidelities.
That is, while realizing fast gates with a nonlinear qubit-motion interaction is generally a challenging task, there are benefits in exploring this regime beyond the obvious advantages of fast gates and the benefit of weaker required driving that comes with stronger qubit-motion coupling.

\begin{figure}
    \centering
    \includegraphics[width=\columnwidth]{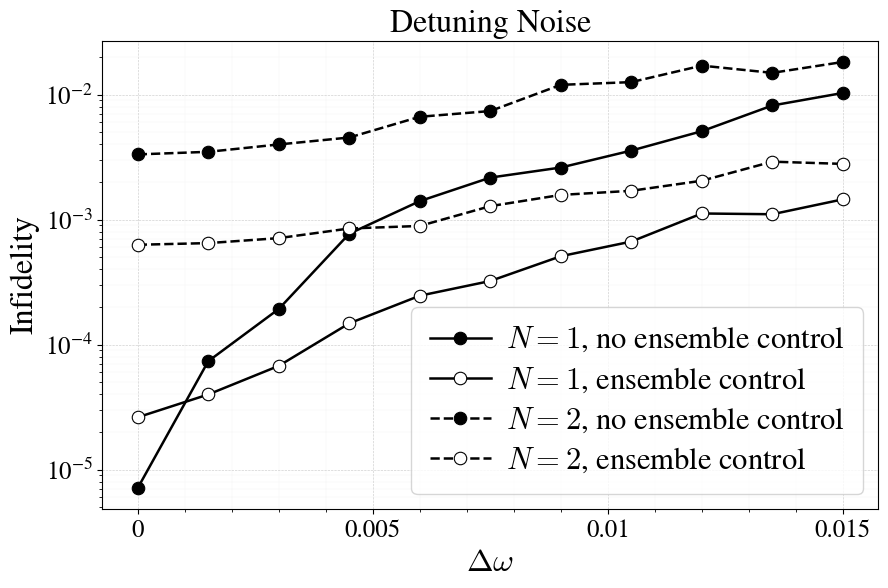}
    \caption{Ensemble-averaged infidelity (Eq.~\eqref{eq:ensemblefid}) with $\bar{n}_1=0$ and $\Delta\epsilon=0$ when subject to detuning noise with $\varepsilon_{ij}$ chosen from $[-\Delta\omega,\Delta\omega]$. The LD parameter is $\eta=0.4$ and the gate duration is 3 trap periods. The hollow/filled points correspond to gates optimized with/without ensemble control and the solid/dashed lines correspond to gates resulting from optimization targets (Eq.~\eqref{eq:Utarget}) with the projector specifying one/two lowest energy Fock states. The plot demonstrates that ensemble control can reduce infidelity, but also that using a larger projector for the motional states can increase the infidelity. The lines connecting data points are a guide to the eye.}
    \label{fig:frequency_noise}
\end{figure}
                        
\begin{figure}
    \centering
    \includegraphics[width=\columnwidth]{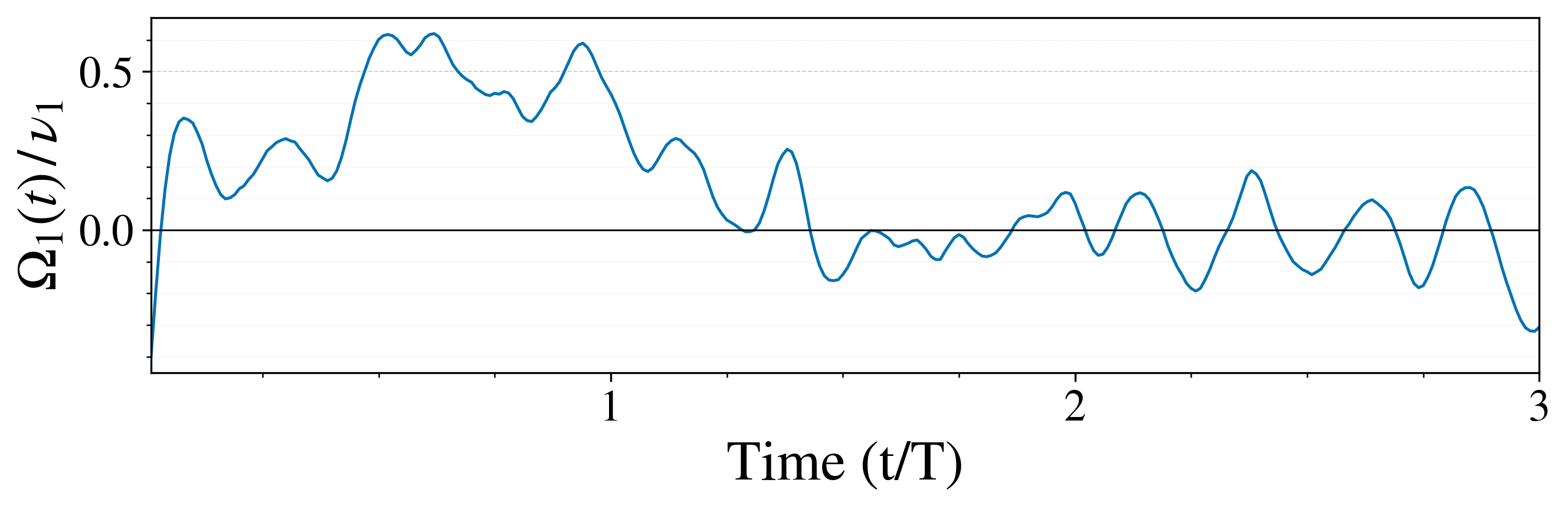}
    \caption{Example $\Omega_1$ for a drive optimized with ensemble control and with a motional projector $P_M$ (Eq.~\eqref{eq:Utarget}) specifying the two lowest motional states; the LD parameter is $\eta=0.4$ and the duration is 3 trap periods. The Rabi frequency is given as a multiple of the trap frequency and the time $t$ is given as a multiple of the trap period $T$.}
    \label{fig:pulse}
\end{figure}

With the goal of achieving resilience to fluctuations in the initial motional state, and to fluctuations in Rabi frequency and detuning, there can be a tradeoff between the three different aspects of resilience; i.e., in optimizing a driving pattern for strong resilience to one of these aspects, one might be forced to be content with reduced resilience to one or both of the other aspects. 
In order to assess the extent of this tradeoff, Fig.~\ref{fig:frequency_noise} depicts the dependence of gate infidelities on the magnitude of detuning noise for gates realized with driving patterns optimized with and without using ensemble control.
The gates are optimized using the same motional projector $P_M=\sum_{i=0}^{N-1}|i \rangle\langle i|\otimes\ket{0}\bra{0}$ in the target (Eq.~\eqref{eq:Utarget}).
The solid and dashed lines correspond to gates optimized with $N=1$ and $N=2$ respectively; and the hollow and filled points correspond to gates optimized with and without ensemble control respectively.
In the noiseless case, the infidelities of the gates optimized with $N=1$ are lower than those of the gates which work with multiple initial motional states ($N=2$), i.e., it is harder to achieve a gate which works for multiple motional states. 
As the error level is increased, the infidelities of the gates which use ensemble control are always lower than those without ensemble control, and the infidelities are lower when using lower-dimensional projectors $P_M$, which in turn reduces thermal resilience.

Overall this highlights that the resilience of gates subject to detuning noise can be increased either by using ensemble control or by choosing a reduced degree of thermal resilience. 
This tradeoff between detuning noise resilience and thermal resilience is likely due to the fact that higher $\eta$ and larger temperatures synergize to enhance detrimental motional transitions which are difficult for the pulses to control. On the other hand the detuning noise contributes an additional time dependence which accumulates error over time; shorter gate duration reduces exposure to this noise and larger $\eta$ diminishes the relative prominence of this error compared to the noiseless motional dynamics.

\subsection{Driving pattern}

In addition to the maximum amplitude of driving fields depicted in Fig.~\ref{fig:ampl}, there is also the aspect of spectral width that is of practical relevance.
Generally, numerical optimization of driving patterns parametrized to be piecewise constant can result in wildly oscillating, seemingly noisy patterns.
Since the highest frequencies in such patterns typically exceed all system frequencies, one can smooth the patterns (e.g., by convolution with a Gaussian function) with only a minor reduction in the fidelity of the resultant gates.
Using such smoothed functions as initial condition for a subsequent optimization (and potentially a repetition of these steps) can yield spectrally well-behaved driving patterns for the realization of strongly noise-resilient gates.

An example of such a pattern is depicted in Fig.~\ref{fig:pulse}.
This pattern realizes the gate with $\eta=0.4$ and $T_G=3T$ in Fig.~\ref{fig:allresilience}. 
With 3 trap periods and 300 time bins, the duration of each time interval can be on the order of nanoseconds to tens of nanoseconds, and the dominant frequency is less than 5 times greater than the motional frequency. Thus the spectral demands are comparable to those of previous works \cite{Chen2025,Lishman2020}. 

\section{Discussion}\label{sec:discussion}

Numerical pulse shaping algorithms enable entangling gates to be designed by solving the laser-ion interaction dynamics nonperturbatively. This approach enhances fidelity by mitigating dominant error sources \cite{Schfer2018,Ballance2016,Gaebler2016} across a wide range of the LD parameter, and permits much faster gates, which can reduce cumulative exposure to decoherence.

Nonlinearity in the qubit-motion coupling, traditionally treated as an obstacle to be suppressed, is here a resource; decreasing the duration of gates and increasing the coupling strength can improve the resilience to noise in the detuning between the laser and electronic transition frequencies. 
Gates are also demonstrated to be highly insensitive to temperature and to considerable uncertainty in the laser intensity; but the optimal parameter regimes for high-temperature gates and gates resilient to detuning noise do not overlap, and this tradeoff should inform experimental design. 

The resulting scheme does not require individual addressing of ions and can be implemented with two distinct amplitude-modulation patterns.
The required spectral widths are similar to those of contemporary methods and the required laser intensity follows the same scaling as the ordinary MS gate.
The method benefits naturally from hardware acceleration and advances in gradient-based optimization methods \cite{Lewis2025}; and future work should focus on extending this approach to include additional laser control fields, time-dependent trap squeezing, and localized phonon modes in longer ion chains \cite{Olsacher2020}. This provides a flexible foundation for further improvements in gate speed, fidelity, and robustness.

\section*{Acknowledgments}
This work was supported by the U.K. Engineering and
Physical Sciences Research Council via the EPSRC Hub in
Quantum Computing and Simulation (EP/T001062/1).
\section*{Data Availability}
The software and data that support the findings of this article are openly
available \cite{stefanescu2026software}.
\bibliography{nonperturbative_ref}
\onecolumngrid
\appendix
\setcounter{figure}{0}
\renewcommand{\thefigure}{A\arabic{figure}}
\renewcommand{\theHfigure}{A\arabic{figure}}   

\section{Control Hamiltonian Derivation}

The control Hamiltonian used in the numerics can be derived by starting with the elementary interaction as defined in Eq.~\eqref{eq:elementaryinteration}. The non-interacting Hamiltonian is given by $H_0
= \frac{\omega_0}{2}\sum_{j=1}^2\sigma_z^j+
\sum_{j=1}^2\nu_ja_j^\dagger a_j\,.$ The interaction Hamiltonian is 

\begin{equation}
H_I = \sum_{j=1}^2
\left\{\sigma_x^j\left[\Omega_1 \cos(kx_j-\omega_1 t)-\Omega_2 \sin(kx_j-\omega_2 t)\right]+
\sigma_x^j\left[\Omega_1 \cos(-kx_j-\omega_1 t)+\Omega_2 \sin(-kx_j-\omega_2 t)\right]\right\}\,,
\label{eq:fullinteraction}\end{equation}
where the sign in front of the $k$ indicates the direction of the traveling field and the $\Omega_2$ on the field traveling in the $-k$ direction has a minus sign relative to the $\Omega_2$ in the $k$ direction. The transformation into the frame defined by $\frac{\omega_0}{2}\sum_{j=1}^2\sigma_z^j$ dresses the Pauli operators with time-dependent phases: $\sigma_+\rightarrow\sigma_+e^{i\omega_0 t}\,.$ The terms written in exponential form have a time dependence like $e^{i(\omega_i\pm\omega_0)t}\,.$ A detuning between the laser and the qubit resonance frequency is defined as $\Delta_i=\omega_i-\omega_0$ and terms with $\omega_i+\omega_0$ are neglected. Both ions interact with these fields in the same way, so all interactions can be expressed as a sum over the ions: 

\begin{equation}
    H = \sum_{j=1}^2\nu_ja_j^\dagger a_j + \left[\sum_{j=1}^2\left( \sigma_+^j e^{-i \Delta_1 t} + \sigma_-^j e^{i \Delta_1 t} \right)
\Omega_1 e^{ikx_j} + \left( \sigma_+^j e^{-i \Delta_2 t} + \sigma_-^j e^{i \Delta_2 t} \right)
i\Omega_2 e^{ikx_j} + \mathrm{h.c.}\right]\,.
\label{eq:fullH}\end{equation}
The noiseless model assumes $\Delta_1=\Delta_2=\Delta$, and choosing $\Delta=0$ reduces the Hamiltonian to 

\begin{equation}
H = \sum_{j=1}^2\nu_ja_j^\dagger a_j + \left[\sum_{j=1}^2\Omega_R\sigma_x^je^{ikx_j} + \mathrm{h.c.}\right]\,.
\end{equation}
To model noise, the substitution is made 
\begin{align}
\Omega_{ij}&\rightarrow(1+\epsilon_{ij})\Omega_j^c\,,
\\
\Delta_{ij}&\rightarrow\Delta_j^c + \varepsilon_{ij}\omega_T\,,
\end{align}
to represent some small unknown perturbations $\epsilon$ and $\varepsilon$. The central detuning $\Delta_i$ is set to zero and the full Hamiltonian is then expressed as $H = \sum_{j=1}^2\nu_ja_j^\dagger a_j+H_{\text{noise}}$ and $H_{i,\text{noise}}$ of a single ensemble member is expressed as 

\begin{align}
H_{i,\text{noise}} = (1+\epsilon_{i1})\Omega_1 \sum_{j=1}^{2} \left( e^{-i\varepsilon_{i1}\omega_T t} \sigma_+^j 
+ e^{i\varepsilon_{i1}\omega_T t} \sigma_-^j \right) e^{ikx_j}
+ i(1+\epsilon_{i2})\Omega_2 \sum_{j=1}^{2} \left( e^{-i\varepsilon_{i2}\omega_T t} \sigma_+^j 
+ e^{i\varepsilon_{i2}\omega_T t} \sigma_-^j \right) e^{ikx_j} + \text{h.c.}
\label{eq:fullHnoise}\end{align}

\section{Numerical Method}

The gradient ascent pulse engineering (GRAPE) \cite{Khaneja2005} algorithm is summarized here as well as how it is applied to the laser-ion Hamiltonian. Generally the system would be governed by a Hamiltonian

\begin{equation}
    H(t) = H_0 + \sum_iu_i(t)H_i(t)\,,
\end{equation}
with constant parts $H_0$ and time-dependent parts $\sum_iu_i(t)H_i(t)$,
and the time-dependent functions to be solved for are approximated by piecewise-constant pulses. In the typical applications the term $H_i(t)$ would be time-independent but in this approach (using the interaction picture) it would also have the time dependence from the dressed motional operators. The time-dependent Rabi frequencies are parametrized by piecewise constant functions, within each time interval, $u_i(t)H_i(t)$ is approximated as being constant, and this approximation is accurate as long as the time intervals are sufficiently small. This approximation can also be verified by using the pulses generated by GRAPE and shorter time bins or by checking the dynamics in the frame defined by the free qubits (defined by $\frac{\omega_0}{2}\sum_{j=1}^2\sigma_z^j$) in which $H_i$ is time-independent and $H_0=\sum_{j=1}^2\nu_ja_j^\dagger a_j\,.$

The optimization target for the noiseless case was chosen to be the gate fidelity defined as

\begin{equation}
    F(U,U_Q\otimes P) = \left|\operatorname{tr}[U^\dagger (U_Q\otimes P)]/(4\operatorname{tr}P)\right|^2\,,
\label{eq:fidelity}\end{equation}
where $U$ is the propagator induced by the Hamiltonian with the pulse $\Omega_R(t)$, $U_Q$ is the target in the computational Hilbert space which can be chosen to be a maximally entangling gate such as $\exp(\frac{i\pi}{4}\sigma_x^1\otimes\sigma_x^2)$ and $P$ is the target in the motional space.

Assuming a constant Hamiltonian within each time interval, $H(t)=H_k$ for $t\in[(k-1)\Delta t,k\Delta t]\,,$ the propagator within each time interval can be easily computed by the first term of the Magnus expansion: $U_k=e^{-i\Delta tH_k}$ and the full time evolution is given by $U_{total}=\prod_{k=M}^{1}U_k\,.$ The target to be optimized is the fidelity (Eq.~\eqref{eq:fidelity}) and the optimal pulses are found by gradient-based optimization. The gradient of the fidelity can be provided to an appropriate optimizer, where the gradient of the propagator with respect to the amplitude of a pulse in any given time interval is given by 

\begin{equation}
    \frac{\partial U_{total}}{\partial u_{ik}}=U_MU_{M-1}...\frac{\partial U_{k}}{\partial u_{ik}}...U_2U_1\,,
\end{equation}
and the gradient of the propagator for each time interval is computed up to second order in $\Delta t\,:$
\begin{equation}
    \frac{\partial U_{k}}{\partial u_{ik}}\approx\left(-i\Delta t\frac{\partial H_k}{\partial u_{ik}}-\frac{1}{2}\Delta t^2\left[H_k,\frac{\partial H_k}{\partial u_{ik}}\right] \right)U_k\,.
\end{equation}

The pulses calculated typically have high spectral width, and smoother solutions may be desired. This can be done in several ways; the approach used in this work is to run an optimizer with a random initial guess and then use a Gaussian filter to smooth the solution and then pass it to the optimizer again for further re-optimization. This can cause the fidelity to drop before re-optimization; several cycles of smoothing and re-optimization may be necessary before a desired balance of smoothness and fidelity is achieved. 

The sensitivity of the resulting dynamics to the temperature as well as laser-qubit detuning and Rabi frequency is relevant to experiments because it is never possible to select these quantities with zero uncertainty nor is it possible to prepare the pure ground state without any thermal impurity. To mitigate the temperature sensitivity, the motional target would ideally be the identity. This is numerically infeasible, so instead a projector is used:
\begin{equation}P=\sum_{{\bf n}}^{N}|{\bf n} \rangle\langle {\bf n}|\,,\label{eq:projector}\end{equation}
where $|{\bf n}\rangle=|n_1,n_2\rangle$ specifies the states of both the COM mode and the relative mode, and $N$ is much smaller than the full size of the truncated bosonic Hilbert space used in the numerical simulation. For this work, $P = |0\rangle\langle0|\otimes|0\rangle\langle0|$ and $P = (|0\rangle\langle0|+|1\rangle\langle1|)\otimes|0\rangle\langle0|$ are used.

The sensitivity of the gate to uncertain quantities can be mitigated via ensemble control \cite{Li2009}, where an average gate fidelity $\frac{1}{N_c}\sum_jF(U_j,U_Q\otimes P)$ or state fidelity $\frac{1}{N_c}\sum_j|\bra{\psi_j}\ket{\psi_{T,j}}|^2$ of an ensemble of systems is maximized and each member $U_j$ or $\psi_j$ in the ensemble is induced by a Hamiltonian $H_j$ in which the unknown perturbing quantities are chosen from some probability distribution.

For enhanced numerical efficiency, state fidelity can be used instead of gate fidelity as an optimization objective function. Since the qubit Hilbert space of the gate includes only 4 basis-state transfers, taking just one of these was found to be sufficient in practice to enforce that the entire gate is realized, i.e. the full gate fidelity must be used for verification after the optimization. In this case, the thermal resilience can also be included via the ensemble control approach, where the average state fidelity over several initial motional states is used. The objective function is given by

\begin{equation}
    \mathcal{F} = \frac{1}{NN_c} \sum_j^{N_c}\sum_{{\bf n}}^{N} \left| \langle \psi_{T,j,{\bf n}}| \psi_{j,{\bf n}} \rangle \right|^2\,,
\label{eq:statefidelity}\end{equation}
with $\ket{\psi_{j,{\bf n}}}=U_j\ket{00}\otimes\ket{n_1,n_2}$ and $\ket{\psi_{T,j,{\bf n}}}=\exp(\frac{i\pi}{4}\sigma_x^1\otimes\sigma_x^2)\ket{00}\otimes\ket{n_1,n_2}\,.$ The full gate fidelity as defined in Eq.~\eqref{eq:gatefidelity} should be used after the optimization to verify that the full propagator is realized with high fidelity.

The bosonic ladder operators are approximated as finite size matrices, ensuring that the sum of the thermal state probabilities $\sum_{n_1,n_2}P_\beta(n_1,n_2) \geq0.9999\,.$ All fidelities reported were calculated with larger bosonic ladder operators than what is used in the optimization; this confirms that leakage to higher energy Fock states has a negligible effect on the dynamics.

\section{Amplitude Dependence on $\eta$}

The regular MS Hamiltonian is linear in the Rabi frequency and LD parameter; and for a given gate speed these two quantities are inversely proportional to each other, i.e., $\eta \Omega\propto \text{const}$. Fig.~\ref{fig:ampleta} shows that this is also true of gates based on the full nonperturbative laser-ion interaction. Despite the fact that this Hamiltonian depends on $\eta$ as $e^{i\eta(a+a^\dagger)}$, the Rabi frequency required to implement gates still depends on $\eta$ as $\Omega \eta \propto \mathrm{const}$.

\begin{figure}[H]
    \centering
    \includegraphics[width=0.5\textwidth]{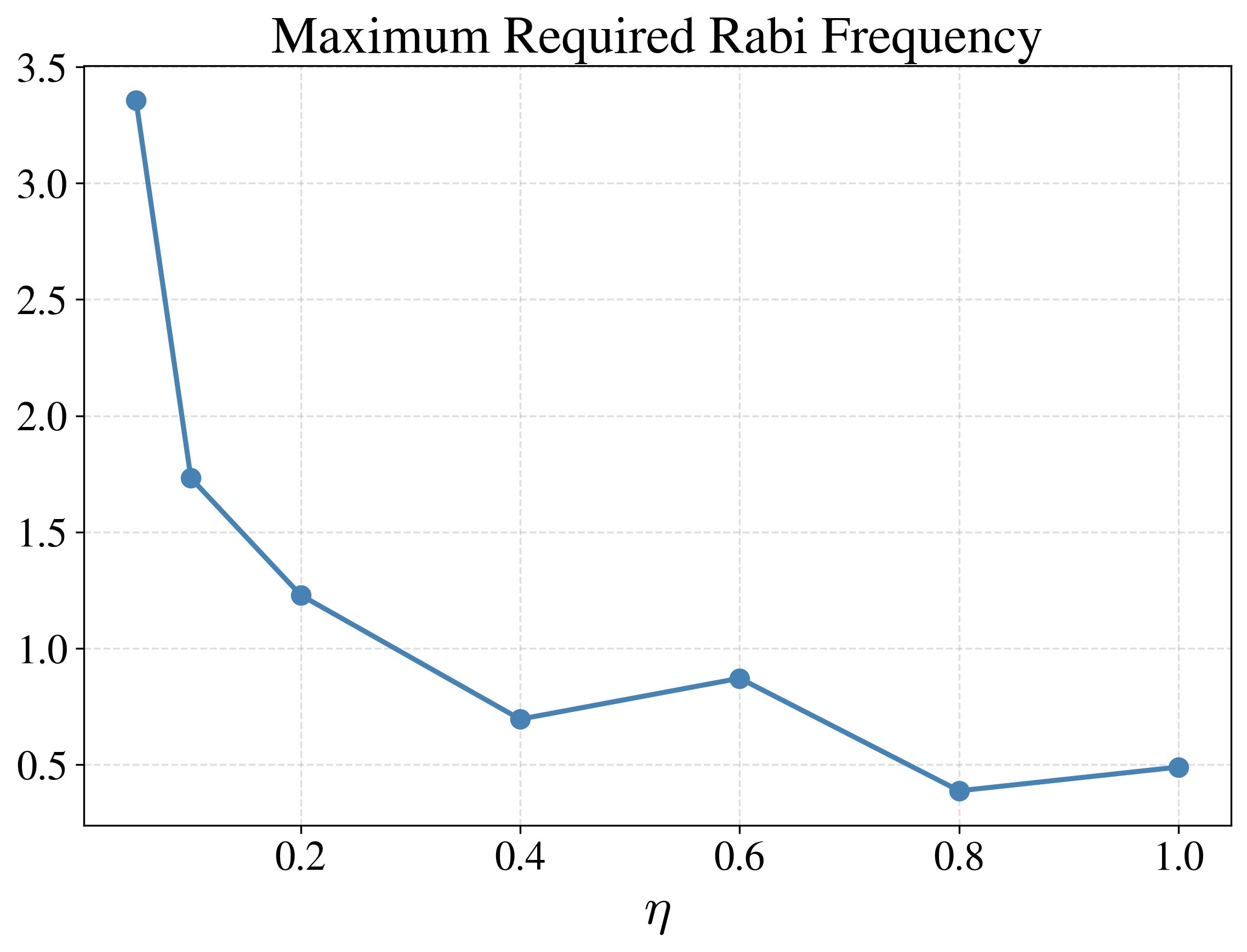}
    \caption{Maximum Rabi frequency reached by the drives for gates designed for 3 trap periods for different values of $\eta$. The lines connecting data points are a guide to the eye.}
    \label{fig:ampleta}
\end{figure}

\section{Temperature Resilience Tradeoff}

\begin{figure}[H]
    \centering
    \includegraphics[width=0.5\textwidth]{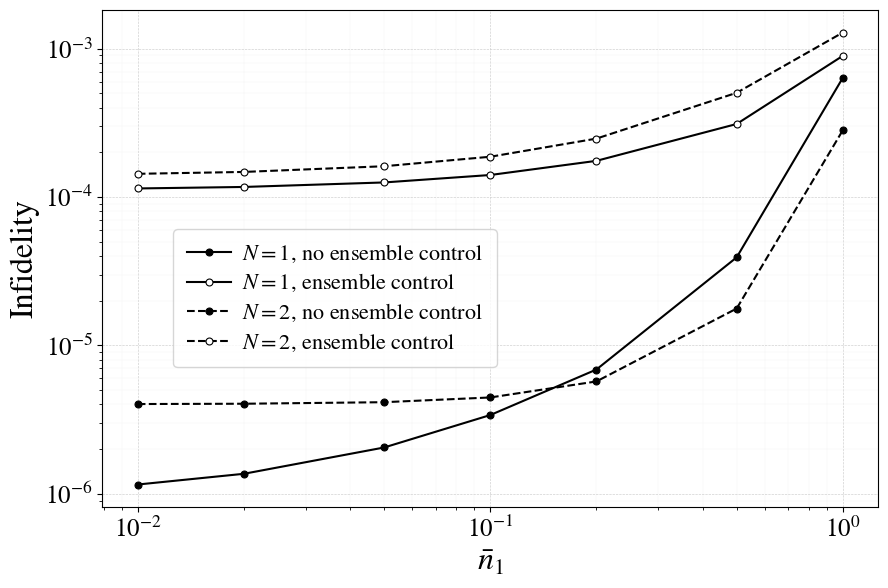}
    \caption{Thermal infidelities of gates with $\eta=0.05$ over 3 trap periods obtained with optimization with an objective function (Eq.~\eqref{eq:Utarget}) targeting the 2 lowest energy Fock states (dashed lines) compared to targeting just the ground state (solid lines). Hollow and filled points correspond to pulses prepared with and without ensemble control respectively, showing the reduced thermal infidelity resulting from forgoing ensemble control. The lines connecting data points are a guide to the eye.}
    \label{fig:projector2}
\end{figure}

Fig.~\ref{fig:projector2} shows the thermal infidelities for gates optimized with and without ensemble control, targeting either one or two motional states. The plot shows that in the regime of fast gates with weak coupling the two gates which use ensemble control achieve roughly the same infidelity regardless of how many motional states are targeted, and the same can be said of gates which do not use ensemble control. However, for gates targeting either one or two motional states, the use of ensemble control increases the infidelity by more than an order of magnitude.
This shows that in this regime, thermal resilience does not need to be explicitly engineered during the optimization but that there is a direct tradeoff when using ensemble control to achieve resilience to detuning noise. 
The ensemble control was also observed to be detrimental to the thermal resilience in the strong coupling regime, although to a lesser extent.

To explain why the degree of thermal resilience achieved in this regime is weakly dependent on the $P_M$ used in the target (Eq.~\eqref{eq:Utarget}), the pure gate infidelity as defined in Eq.~\eqref{eq:gatefidelity} is plotted as a function of different initial motional states. The gates in Fig.~\ref{fig:purethermal} are optimized using $P_M=|0 \rangle\langle 0|\otimes\ket{0}\bra{0} + |1 \rangle\langle 1|\otimes\ket{0}\bra{0}$ in Eq.~\eqref{eq:Utarget}, i.e., they are designed to work for the two lowest energy states. What the figure shows is that when $\eta$ is small, the gate still functions for a large range of states beyond what is explicitly targeted during the optimization. This behavior is similar to that of the MS gate even though the carrier transitions must be accounted for, whereas the MS gate fully neglects them. So although an analytically exact solution for the propagator is not known in this regime, these simulations are a strong indication that it is weakly dependent on the motional state.

Because the optimized propagator naturally retains high fidelity for a few Fock states beyond the targeted subspace; and
because low temperature thermal mixtures concentrate most of the population in the ground state of the COM mode and the unitary dynamics mostly takes place in the COM mode, only two low energy states in this mode need to be explicitly targeted during optimization in order to deliver the thermal resilience reported in the main results of the paper.  

Fig.~\ref{fig:purethermal} shows that this effect does not hold when $\eta$ is large. 
In this case the resulting infidelity clearly reflects that the gate was designed to work only for the two lowest energy motional states. In this regime, explicitly specifying additional motional states that the gate should be optimized for (via $P_M$) reduces infidelity when the gate is applied to thermally mixed motional states.

\begin{figure}[H]
    \centering
    \includegraphics[width=0.5\textwidth]{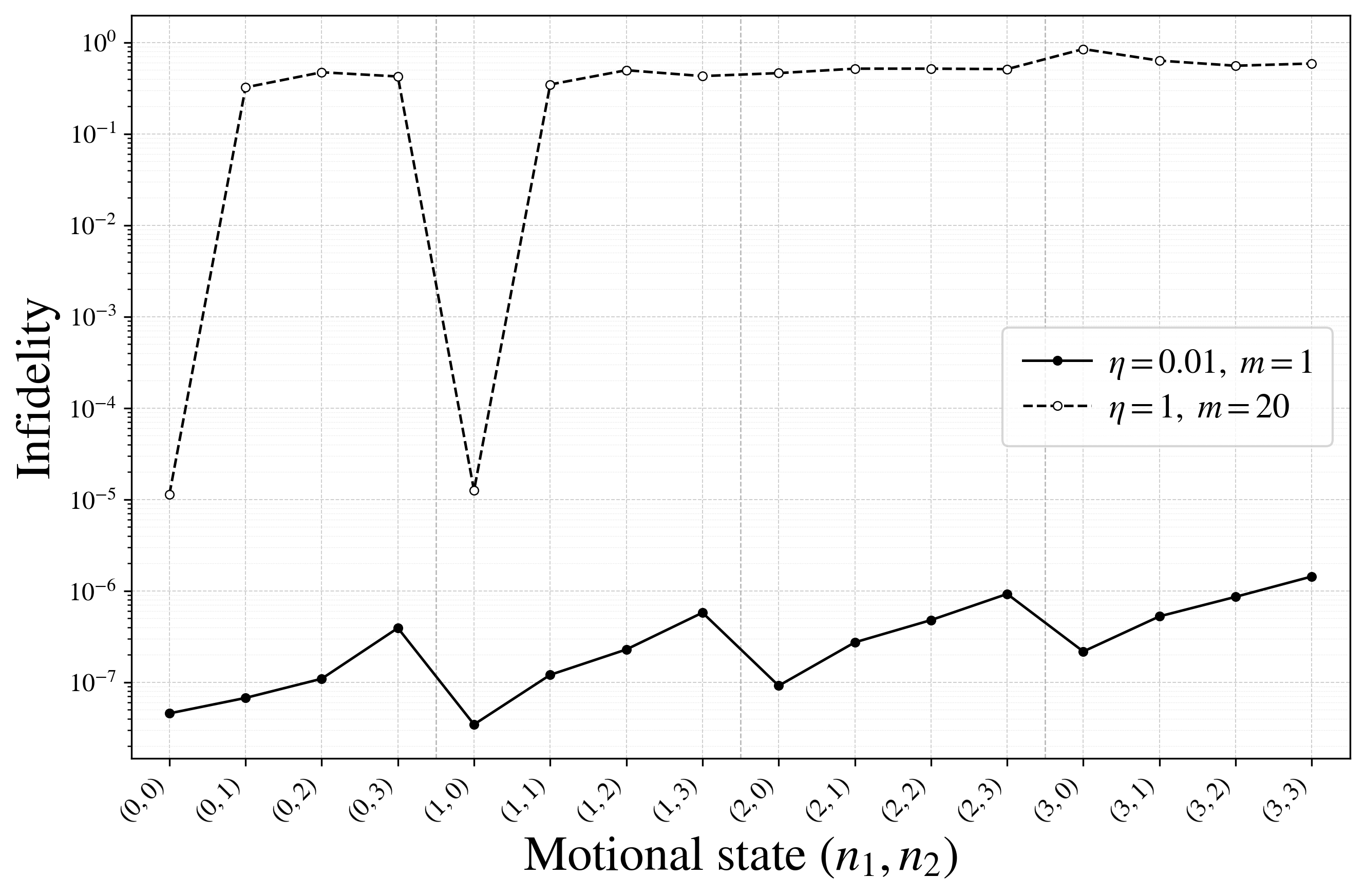}
    \caption{Infidelity as defined in Eq.~\eqref{eq:gatefidelity} vs initial motional state of the COM and relative modes ($n_1,n_2$) for two gates over durations of $m$ trap periods. The lines connecting data points are a guide to the eye.}
    \label{fig:purethermal}
\end{figure}

\vspace{1cm}
\end{document}